\documentclass[siggraph]
    {acmarxiv}

\arxivText{my arxivtxt}

\usepackage{relsize}
\usepackage{xspace}

\def\optix{OptiX\xspace}

\def\ahOnly{\texttt{AH-only}\xspace}
\def\chOnly{\texttt{CH-only}\xspace}
\def\stableOrder{\texttt{stable-next}\xspace}
\def\whileWhile{\texttt{while-while}\xspace}
\def\whileMerged{\texttt{while-merged}\xspace}
\def\rejectRepeats{\texttt{reject-repeats}\xspace}

\definecolor{ForestGreen}{HTML}{009B55}

\usepackage{listings}
\title{Ordered Front-to-back Any-Hit Traversal in RTX}
\author{Ingo~Wald}
\email{Orcid~ID~0000-0003-0046-713X }

\definecolor{darkgreen}{rgb}{0,0.5,0}
\definecolor{midgreen}{rgb}{0,0.6,0}
\definecolor{lightgreen}{rgb}{0,0.8,0}
\definecolor{darkred}{rgb}{0.6,0,0}

\def\code#1{{
    {\texttt{#1}}}}

\begin{abstract}
  We look at the problem of \emph{Ordered Front-To-Back Any-Hit Traversal}
  (FTB); i.e., a traversal that iterates through successive hits
  along a ray in a guaranteed front to back-sorted order, and without
  skipping any intersections even if they occur at the same distance.
  We describe multiple different ways of solving this problem within
  the constraints of the existing ray tracing pipeline, and evaluate
  the different realizations.
\end{abstract}

\begin{document}

\maketitle
\section{Introduction}
\label{sec:intro}

Hardware-accelerated ray tracing has significantly changed ray
tracing. However, using this technology requires developers to
express their ray tracer within what we today call the \emph{(RTX) ray
tracing pipeline}, in which programs are expressed as a combination of
user-programmable ray-gen, any-hit (AH) and closest-hit (CH) programs,
and which forms the common abstraction behind each one of the three
main hardware ray tracing APIs (i.e., OptiX~\cite{optix},
Vulkan~\cite{vkr-spec}, and DirectX~\cite{dxr-spec}). 

For most applications, re-formulating them through AH and CH programs
is straightforward: the next closest surface can
be found with CH, and iterating over all hits along a ray
can be done with an AH
program that simply processes but then rejects each hit. However,
some applications require
iterating over multiple intersections---or \emph{hits}---along a given ray
in a way that
visits them in a guaranteed front-to-back order, while also guaranteeing
 that it will visit every hit if some happen at
the same distance. In graphics, typical examples of that are scenes
with lots of semi-transparent surfaces, decals, or applications
such as ray tracing
of Gaussian Splats~\cite{gaussian-rt}. Outside of graphics the same
problem often arises in simulations where rays have to penetrate
through what are logically solid objects, but which are each
represented through their boundary surfaces (which
become co-planar where such solids abut). Throughout this paper, we
refer to this problem as the ``Front-To-Back Any-Hit'' problem (or
FTB, for short): like AH programs we want to be able to iterate over
possibly all hits along the ray (without missing any), but unlike AH
programs we need hits to be reported in a strictly front-to-back
order.

At first glance, this operation looks like a straightforward
combination of AH and CH, but it isn't: AH programs can be trivially
used to find all hits, but not in order; and CH programs can trivially
be used to find the respectively next closest intersection, but cannot
find more than one hit at any given distance.  In this paper, we
describe different ways of realizing this kernel within
the hardware ray tracing pipeline; i.e., through 
combinations of any-hit and closest-hit programs. We describe,
classify some algorithmic properties of, and evaluate the performance
of those kernels, allowing the reader to choose which is best for a
given situation.

\section{Background and Related Work}

The key problem we want to solve is to trace rays such that are 
guaranteed to find all intersections along that ray---even if they are
co-planar---while also reporting those in ascending order. To
understand why this is tricky---as well as how exactly our proposed
techniques will eventually work---we need to briefly review some of
the more lower-level properties of the RTX Pipeline. We assume basic
familiar with the RTX Pipeline, but some of the techniques and issues
we discuss require a fair amount of detail knowledge that we do not
want to assume.

\subsection{Floating-Point Numbers}

Any hardware implementing the RTX pipeline may or may not internally
use different levels of precision, but all values that go into or come
out of that pipeline---like \code{ray.tMin/tMax}, hit distances,
etc---will be represented in IEEE float32 precision.  Floats are often
viewed as the computer equivalent of the space of real numbers
$\mathbf{R}$; but for many of the techniques in this paper it is
important to remember that \code{float}s are not continuous, and can
only represent a finite, discrete subset of $\mathbf{R}$. For any
``regular'' float $f$ (i.e., any float that is not +/-\code{INF},
+/-\code{FLT\_MAX}, \code{NAN}, or a denorm) there are well defined
next larger respectively next lower values that in this paper we will
refer to as \code{justAbove(f)} and \code{justBelow(f)}. In
C/C++/CUDA, these can be computed using \code{nextafter(f,t)}.  No
computation can ever return a value that lies \emph{between} \code{f} and
\code{justAbove(f)}, or \code{f} and \code{justBelow(f)},
respectively.

\subsection{Floats, Rounding, and Reproducibility}

Floating point numbers are fully deterministic in  that the
same computations performed in the same order with the same inputs
will always produce the same exact same output---but they are \emph{not}
invariant with respect to mathematically equivalent reformulations.
E.g., $a+(b+c)$ and $(a+b)+c$ are mathematically
equivalent, but in floating point can compute
different results.
Consequently, throughout this paper we only consider kernels that modify
\code{ray.tMin/tMax}, but \emph{never} modify the actual ray origin,
direction, etc.

\subsection{The Ray tMin/tMax Interval}

In the RTX pipeline, rays carry \code{ray.tMin} and \code{ray.tMax}
values that allow for specifying a \emph{valid ray interval} that any
intersection has to be in. For the rest of this paper, understanding
how this interval works---and interacts with AH and CH papers---is
crucially important. We emphasize three details: first, this interval
is \emph{ex}clusive, so \code{tMin} and \code{tMax} are explicitly
\emph{not} valid hit distances---to be considered valid any hit at
distance \code{tHit} has to fulfill \code{ray.tMin < tHit < ray.tMax}.
Second, these tests happen automatically and \emph{before} any
user-provided AH program can see the respective hit. And third, the
\code{ray.tMax} value will \emph{not} always remain the same value
specified when launching the ray, but can shrink during traversal:
every time the pipeline ``accepts'' a hit
(see~\ref{sec:ignore-but-accept}) at distance \code{tHit} this
\code{tHit} will become the new \code{ray.tMax}.

Together, these two facts explain why CH programs struggle with hits
at same distance: once the first gets found and accepted at
\code{tFound}, the ray interval gets shortened to that distance, and
all others at the same distance will get automatically rejected; and
any follow-up ray will either find exactly that same intersection (if
it was launched with \code{tMin<tFound}), or will miss all these
intersection altogether (if launched with \code{tMin>=tFound}). On the
other hand, together with some of the float related properties we
discussed above we can sometimes use these rules to our advantage: if
we trace a follow-up ray with \code{tMin=justBelow(tFound)} we know
that hits at \code{tFound} will be valid, yet no other, closer hits
can possibly be valid. Conversely, any ray traced with
\code{tMin=tFound} will \emph{not} ever report any hits at that
distance, yet no other hit beyond \code{tFound} can ever possibly get
rejected by that \code{tMin} value, either.

\subsection{Cost of Any-Hit Programs}

Using AH programs is expensive. This is obviously true if the program itself
is costly, but it is also true even if that program is
trivially simple. This is because if there were no AH programs at all
the hardware ray tracing cores can simply update the \code{ray.tMax}
value in hardware, and can keep on going until the entire traversal is
done. If there is a AH program, however---no matter how cheap---then
the hardware traversal has to be interrupted after every found
intersection, control has to pass back to a CUDA core to execute the
AH program, then hardware traversal has to be resumed after the AH
program returns, etc; this is expensive even if the AH program does little
actual work. The pipeline does allow to disable AH programs on a
per-ray basis (by passing a \code{OPTIX\_DISABLE\_ANYHIT} flag to the
\code{optixTrace()} call), but as discussed before, only AH programs
can ever see more than one hit at a given distance.

\subsection{Accepting vs Ignoring Hits}
\label{sec:ignore-but-accept}

AH programs can influence traversal in two ways. One
is that they can terminate traversal by calling
\code{optixTerminateRay()}. The ray may still execute the respective
CH program on whatever was already accepted, but calling this function
will immediately stop any further traversal steps, ray-triangle
intersections, or AH calls.

The second
is for the AH program to ask the
pipeline to \emph{ignore} a given intersection
(\code{optixIgnoreIntersection()}). The pipeline's default
behavior----if there is no AH program to be executed, or the user
explicitly disabled AH programs for this ray---is to ``accept'' each
hit, and shorten \code{ray.tMax} to that distance. For some of our
methods we need to tell the pipeline to not do that, typically because
doing so would automatically discard any further intersections at that
distance. This, however, means that some of our kernels will
explicitly (have to) tell the RTX pipeline to \emph{reject} a hit even
though the program's own logic will actually accept that same hit for
its own purposes, and vice versa. These situations of
``OptiX-rejecting while actually accepting'' (and vice versa) can look
like implementation erros---but are actually intentional. Where
appropriate we will use the terms \emph{\optix-accept/-reject} to
indicate that this refers only to the program tells the pipeline.

We also point out that (\optix-)ignoring hits will always come at a
cost: shortening the ray can avoid some future traversal steps,
intersections, and AH call; and not accepting hits will preclude these
savings---so our kernels will always try to accept any hit they
possibly can.  In most cases the ideal situation would be that we can
accept a hit, but shorten \code{ray.tMax} only to
\code{justAbove(tHit)}---but \optix does not currently allow for this
in AH programs.

\subsection{Related Work and Multi-Hit}

The particular problem we are addressing in this paper---front-to-back
any-hit---has not yet received much attention in the academic
literature, but a very similar problem (namely, the so-called
\emph{multi-hit} problem) has (e.g.,~\cite{Gribble2014RayTraversal,Gribble2015AnEO,Gribble2016NodeCM,Gribble2016ImplementingNC,Zellmann2017RayTV}), often
in the context of penetration simulations similar to
Butler et al's \emph{Bullet Ray Vision}~\cite{brv}. The task of
a multi-hit traversal is to find, for a given $N$, the first $N$ hits
along a ray, in sorted order, including hits at same
distance. Clearly, if we have a solution to FTB we have automatically
solved multi-hit (by just calling FTB $N$ times). Similarly, we can
largely (but not entirely) reduce FTB to multi-hit: once we have $N$
hits from multi-hit we can get the first $N$ FTB results from that
list of multi-hit results. 

However, things get interesting if we ever need the \emph{next} $N$
hits beyond what was already found, because then we run into the same
problem with hits at same distance as we do for FTB. For completeness
we also discuss how our techniques can be used to extend multi-hit to
find the respectively next $N$ hits, but this is not the focus
of this paper.

In~\cite{Wald2018RobustIF}, Wald et al.~proposed a method that allowed
for iterating through individual hits by replacing the usual depth
first BVH traversal with a new traversal that maintains depth-sorted
queues of yet to be traversed nodes and yet to be reported hits,
reporting the already found hits to the user only when it can be
guaranteed that no closer one could ever be found. This method is
correct and does not require multi-hit; but it does requires modifying
the underlying ray tracer's BVH traversal, which wouldn't be possible
in a hardware accelerated pipeline.

In rendering it is often possible to use approximate solutions.  For
example, Knoll et al~\cite{Knoll2021} described a method that can
report hits in order while also using RTX for acceleration; but at the
risk of possibly dropping some hits. A similar technique is also used
in Gaussian Ray Tracing~\cite{gaussian-rt}. Where applicable these
techniques will probably be faster; in this paper we only consider
methods that are guaranteed to be correct. All our implementations
will be based on OptiX~\cite{optix}, used through the OptiX
Wrapper Library (OWL)~\cite{owl}; but implementations without OWL,
or using Vulkan or DirectX, should be straightforward.

\begin{figure*}
\begin{centering}
\setlength\tabcolsep{.1ex}
\begin{tabular}{cccccc}
  \hline
  \multicolumn{6}{c}{
    {Simple \code{(N dot D)* primID} style shading of our test models, on first/closest hit along the ray}}
  \\
 \hline
 \includegraphics[width=.165\textwidth]{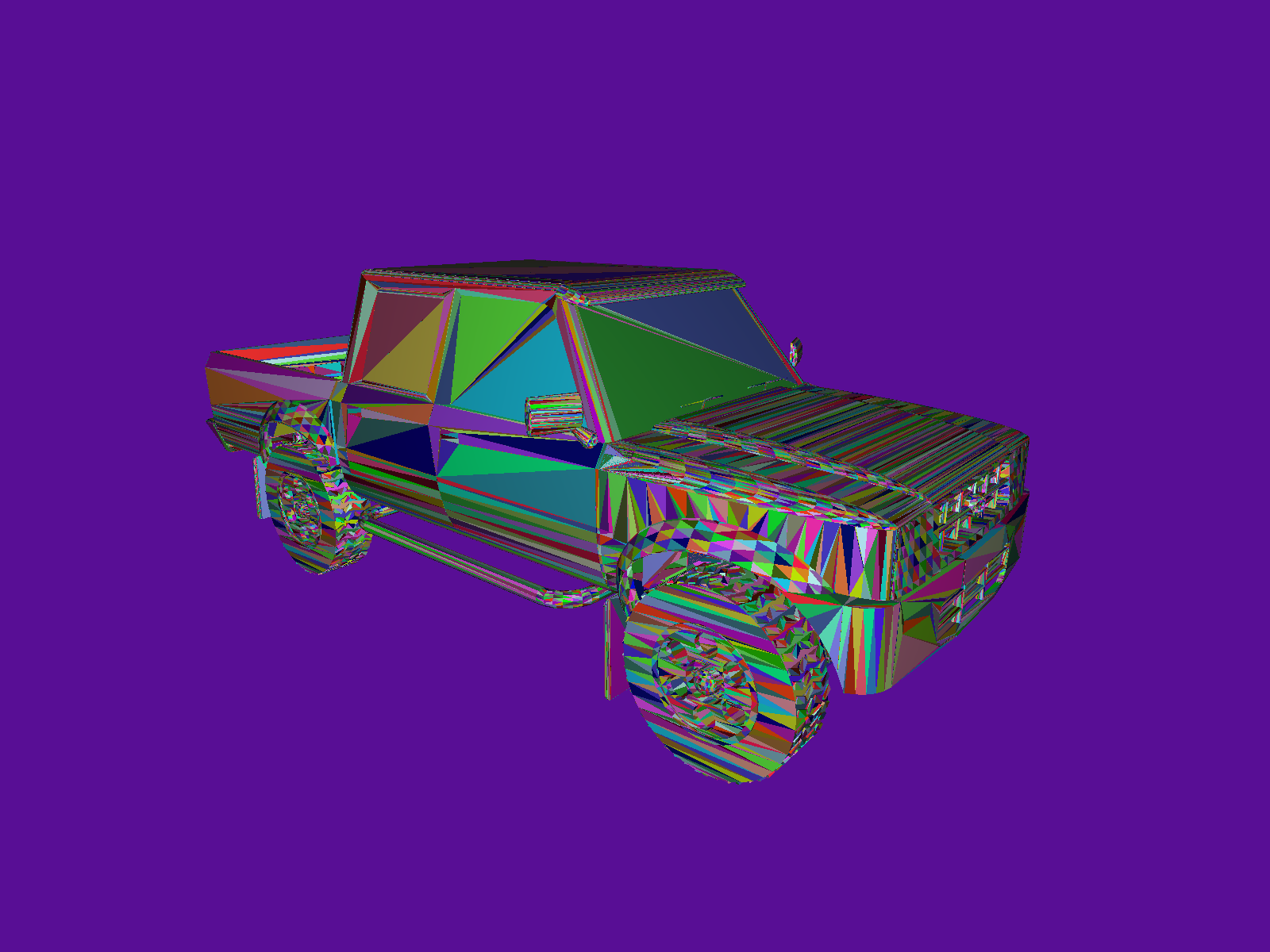}
&
\includegraphics[width=.165\textwidth]{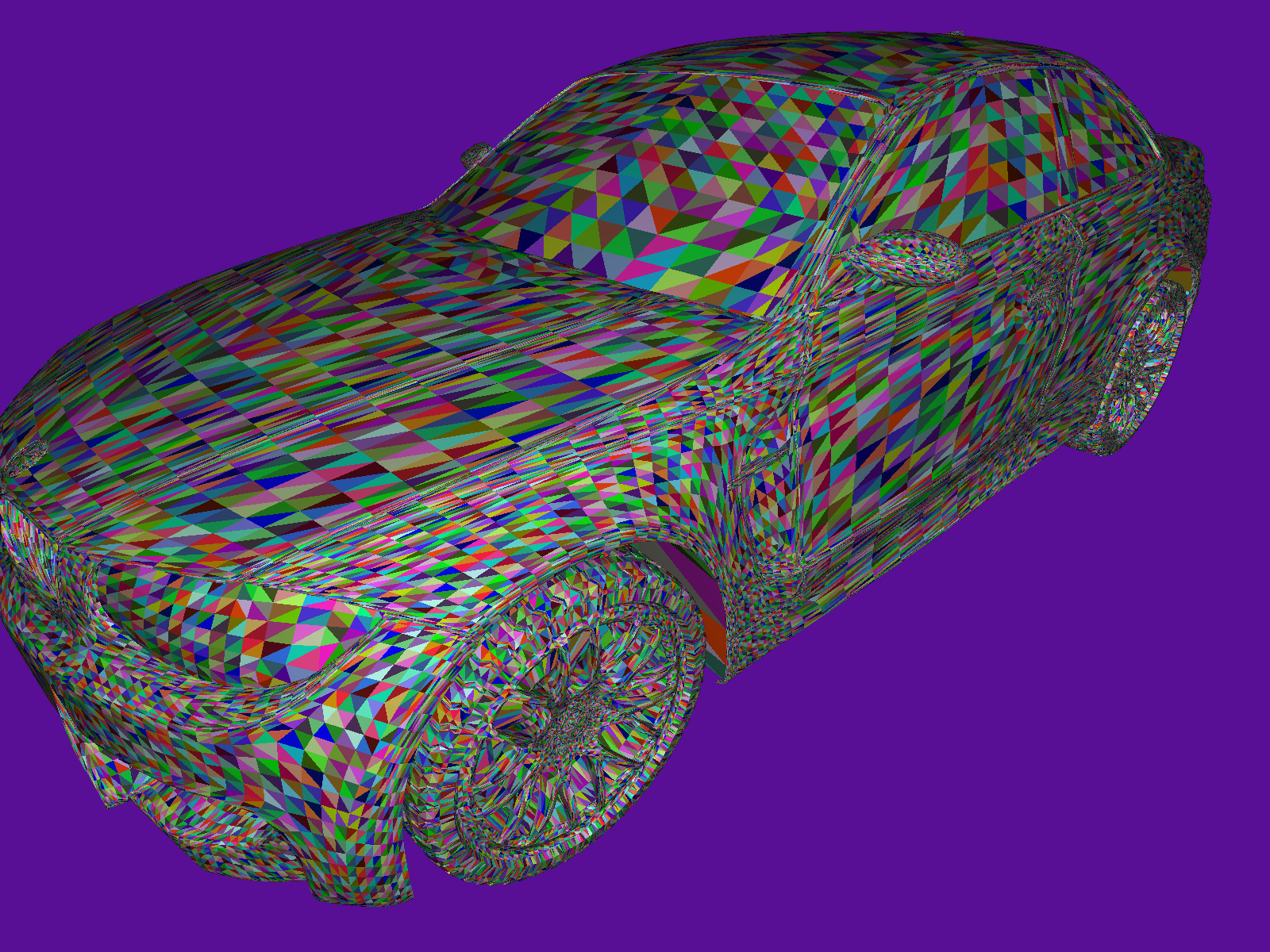}
&
\includegraphics[width=.165\textwidth]{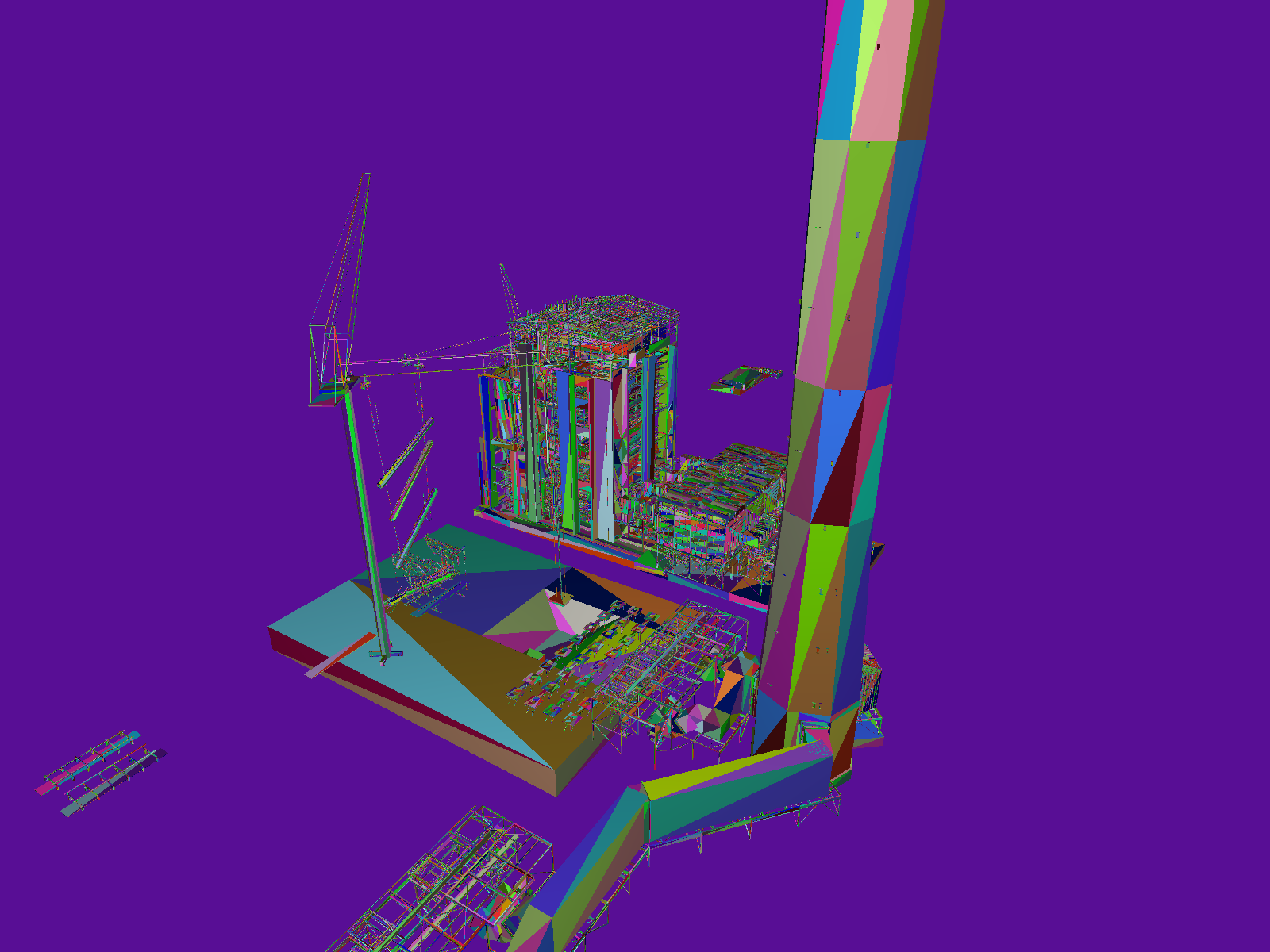}
&
\includegraphics[width=.165\textwidth]{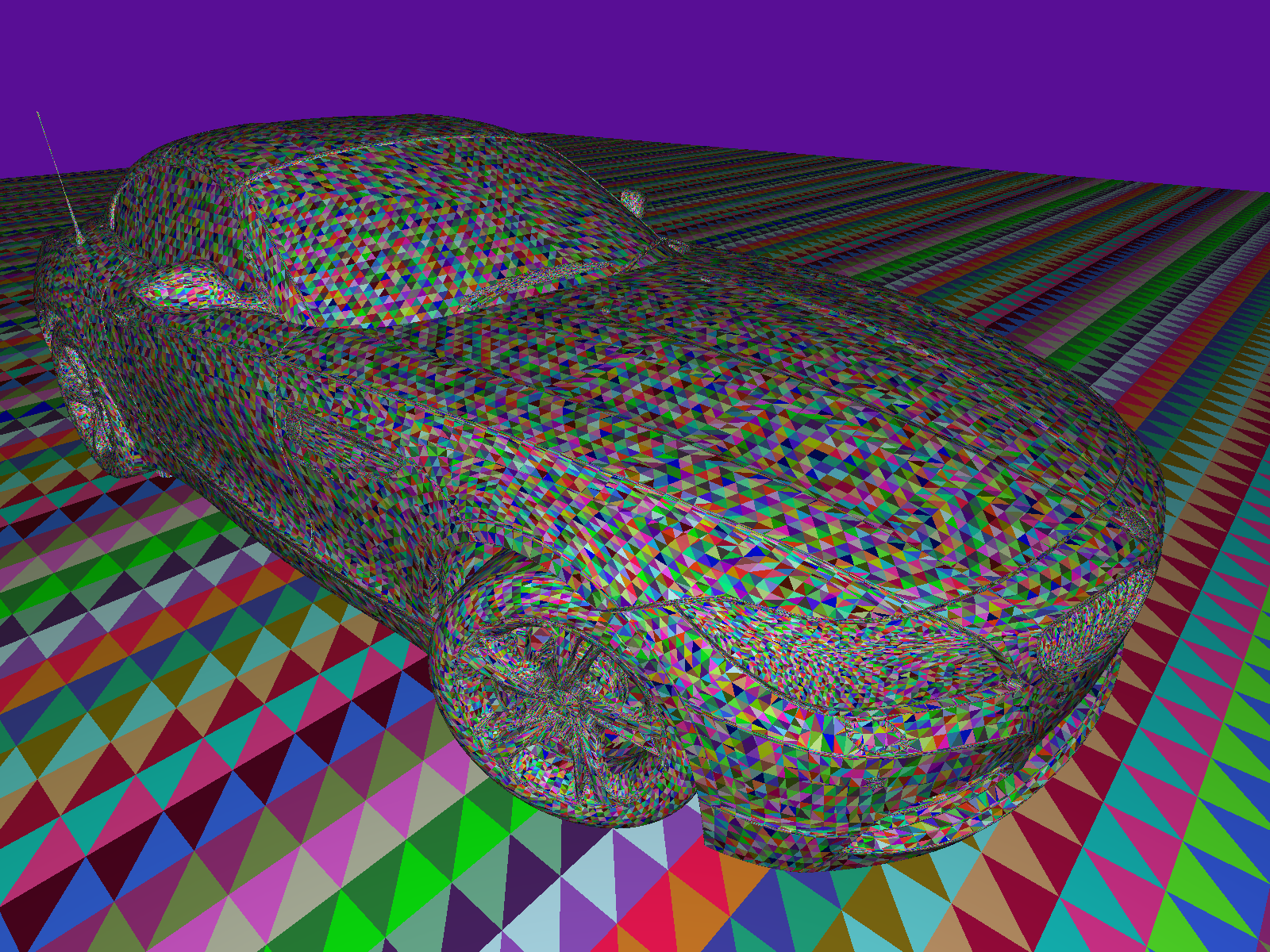}
&
\includegraphics[width=.165\textwidth]{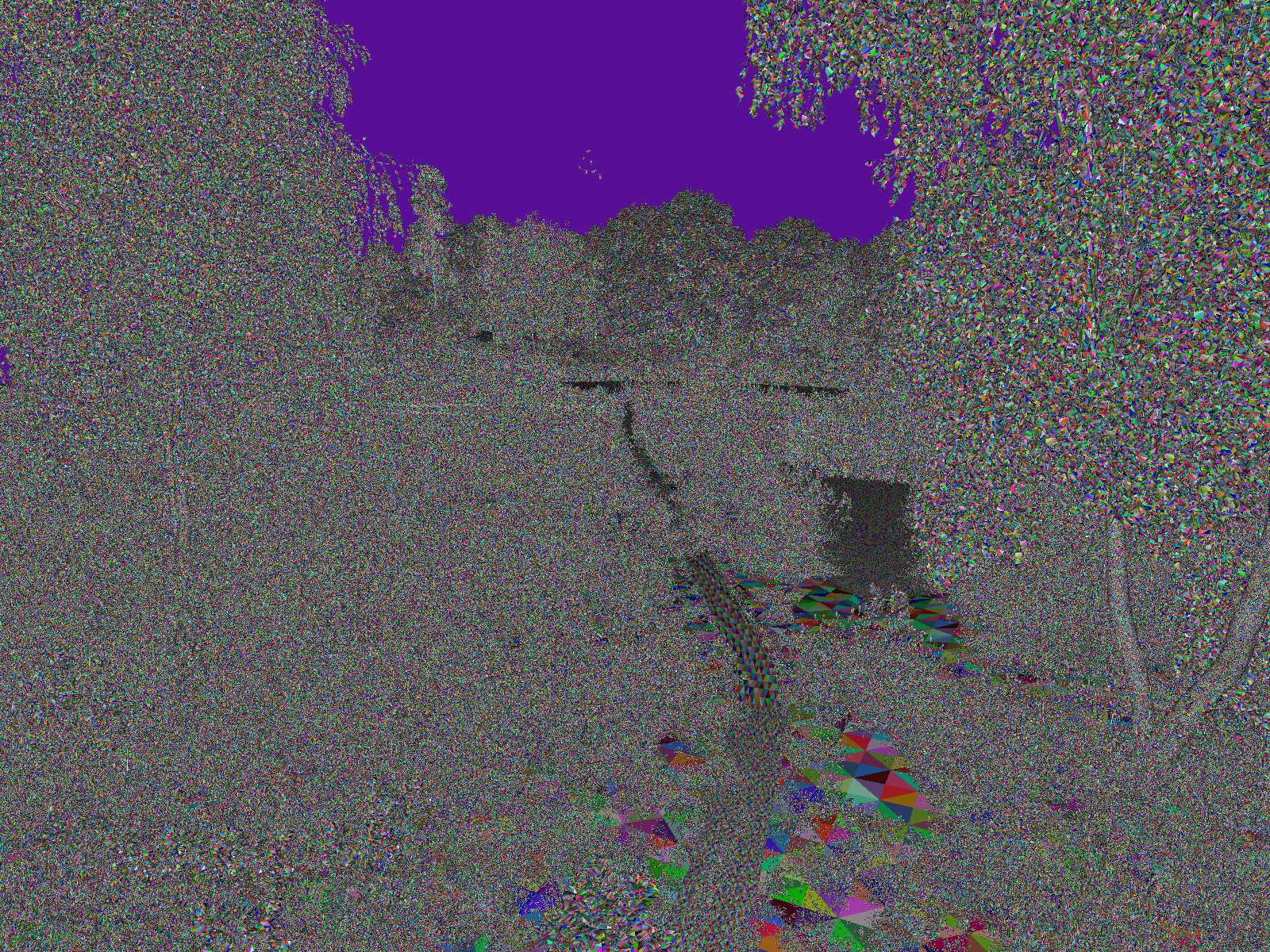}
&
\includegraphics[width=.165\textwidth]{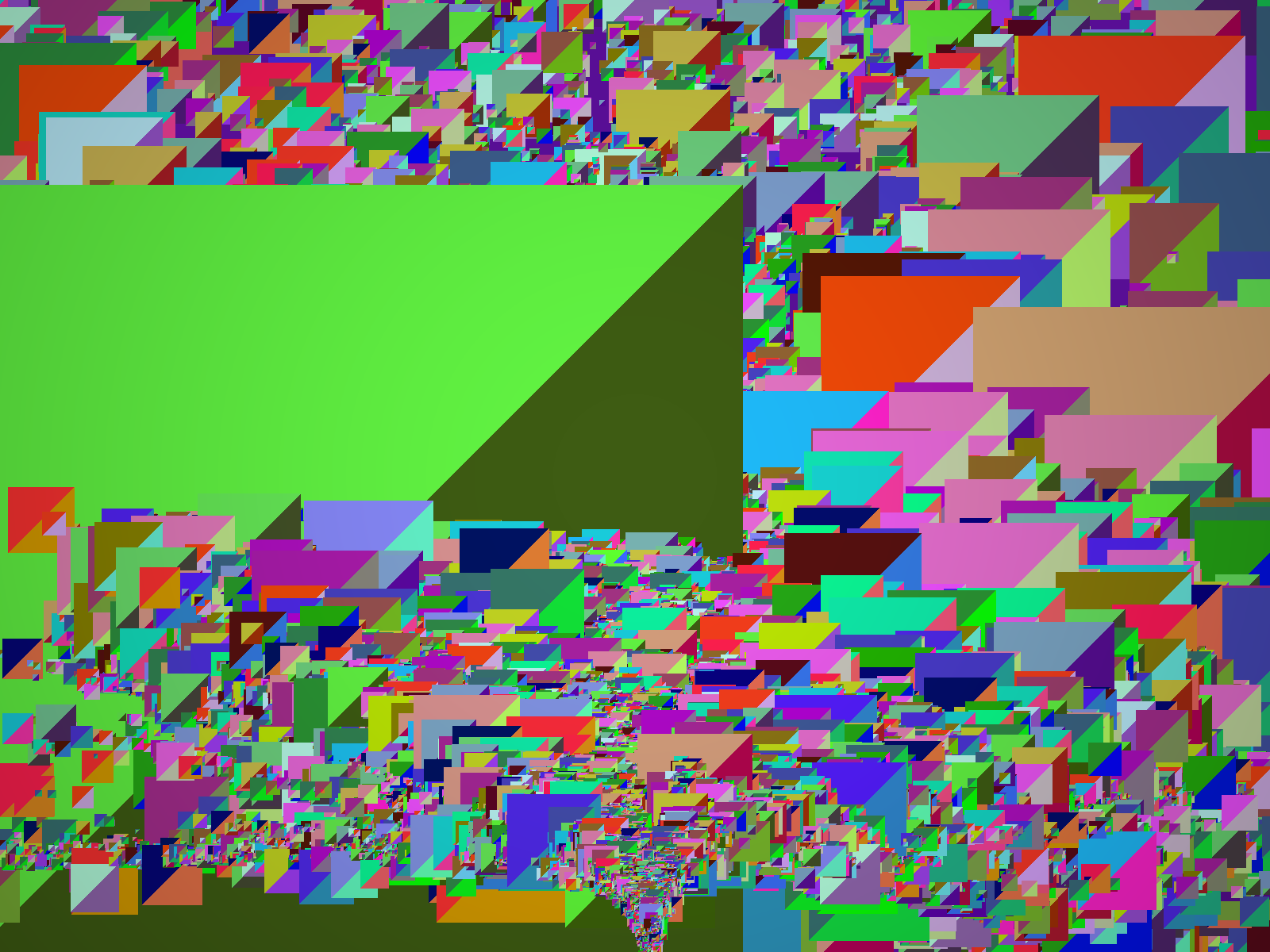}
\\
  \hline
  \multicolumn{6}{c}{
    {pseudo-color shading of number of surfaces encountered when iterating through all intersections}}
  \\
 \hline
\includegraphics[width=.165\textwidth]{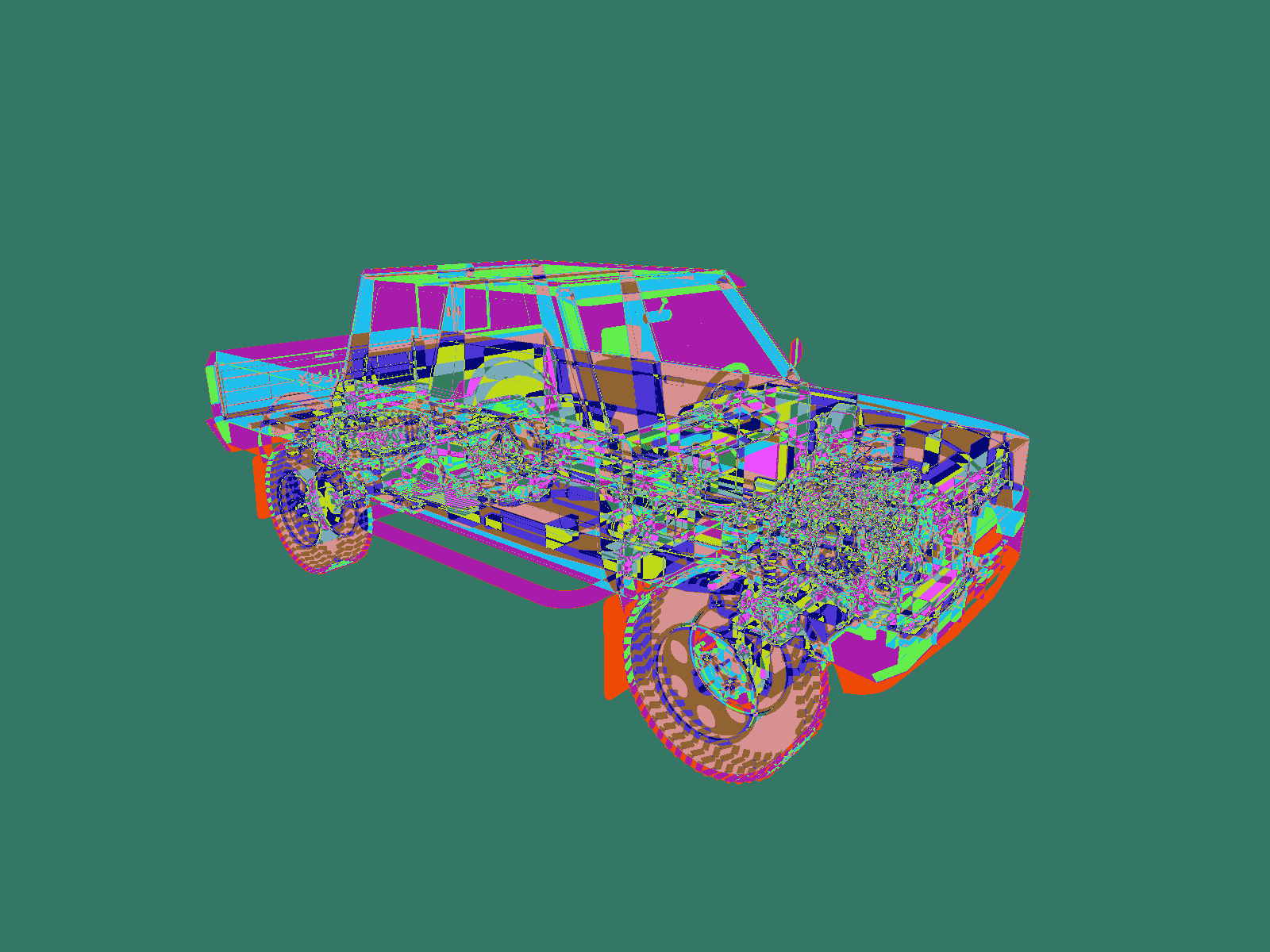}
&
\includegraphics[width=.165\textwidth]{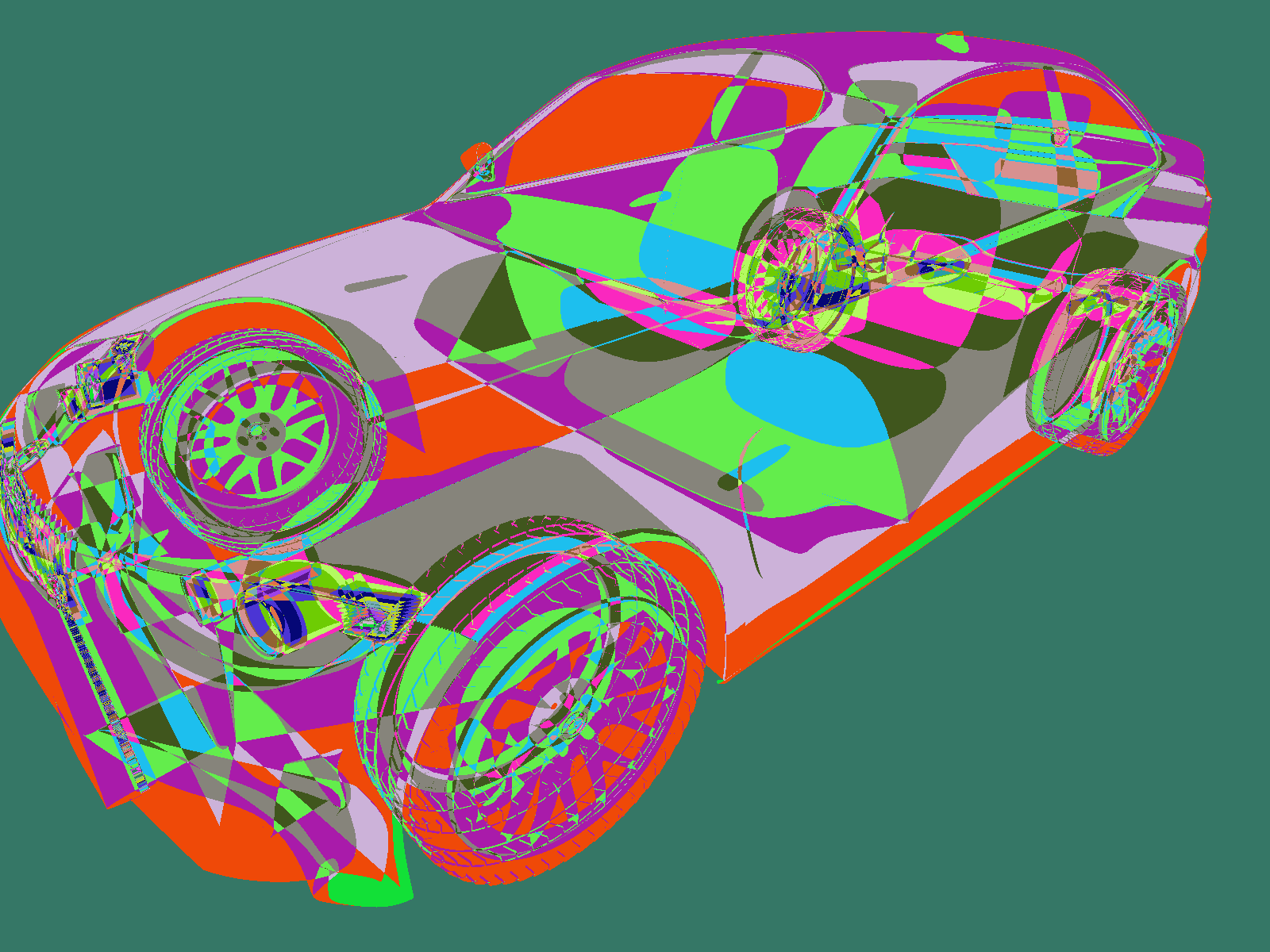}
&
\includegraphics[width=.165\textwidth]{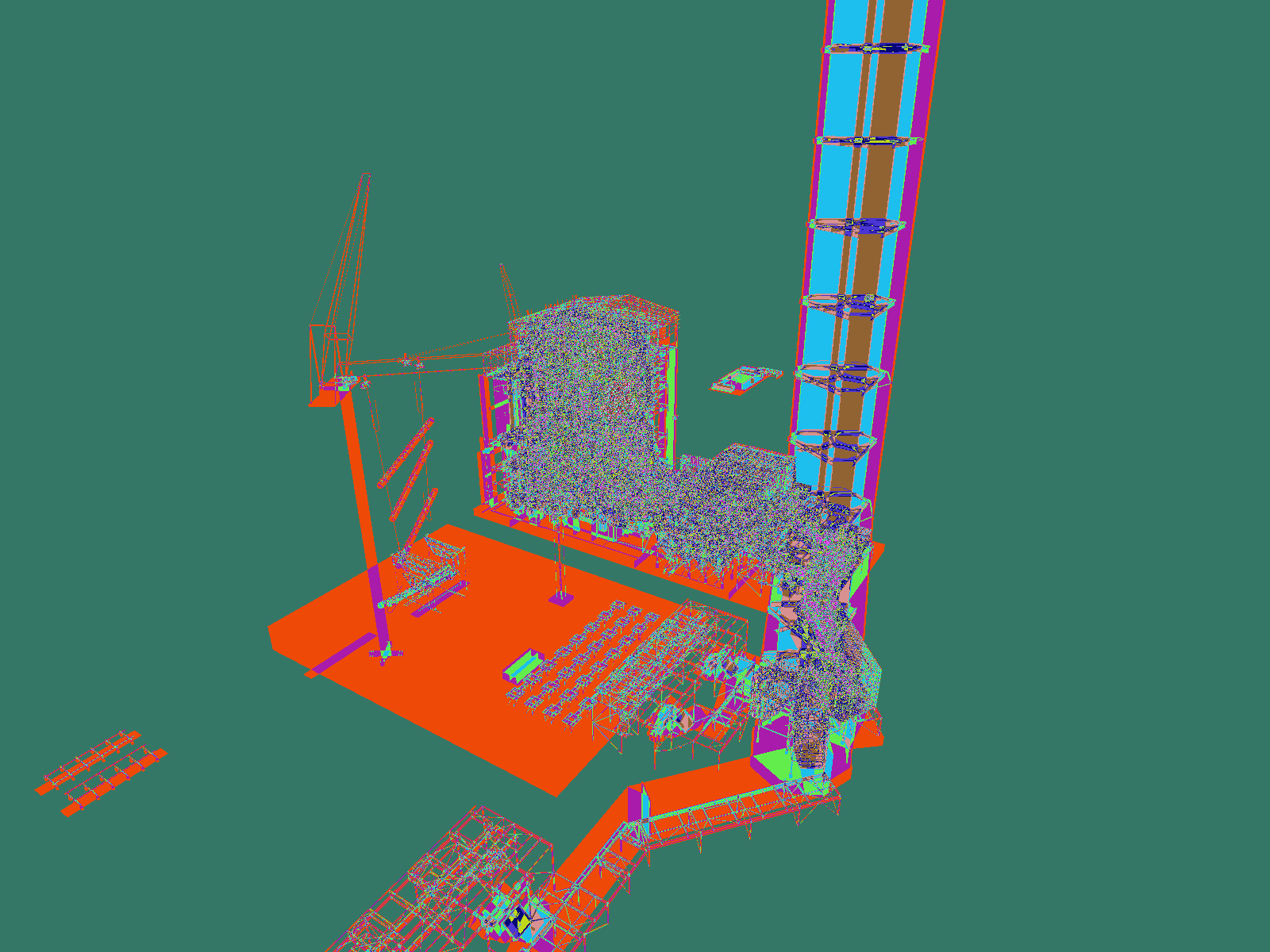}
&
\includegraphics[width=.165\textwidth]{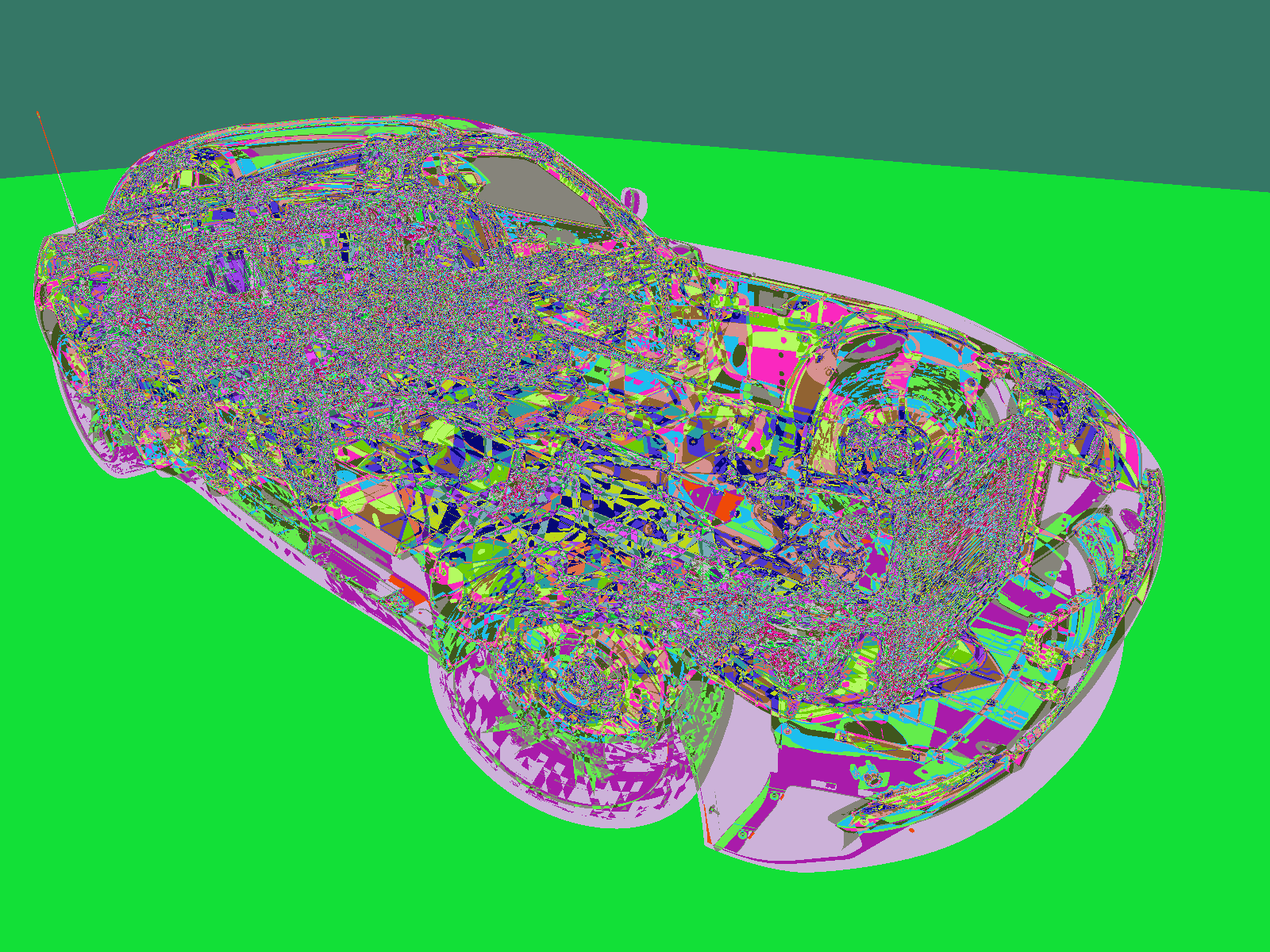}
&
\includegraphics[width=.165\textwidth]{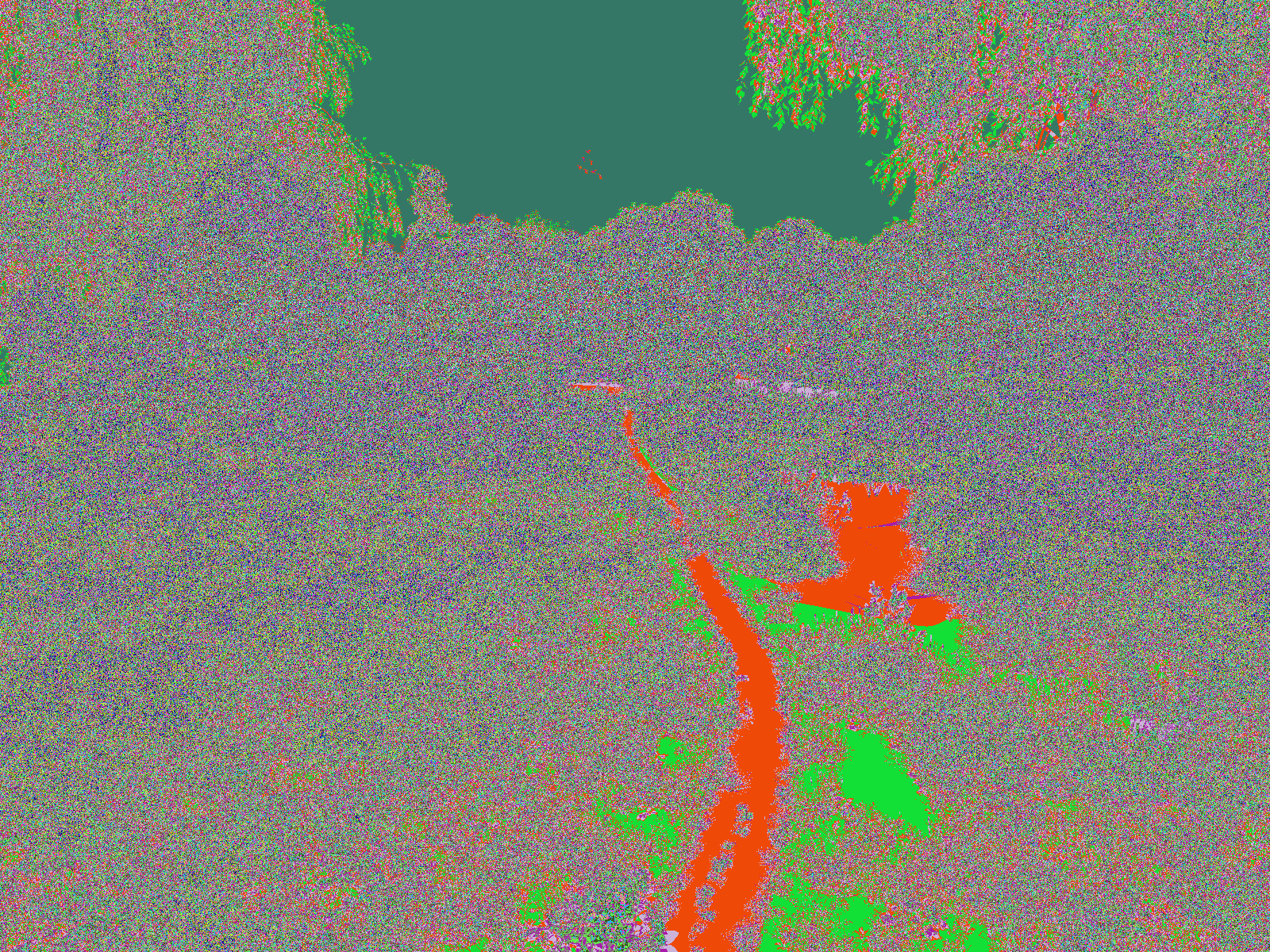}
&
\includegraphics[width=.165\textwidth]{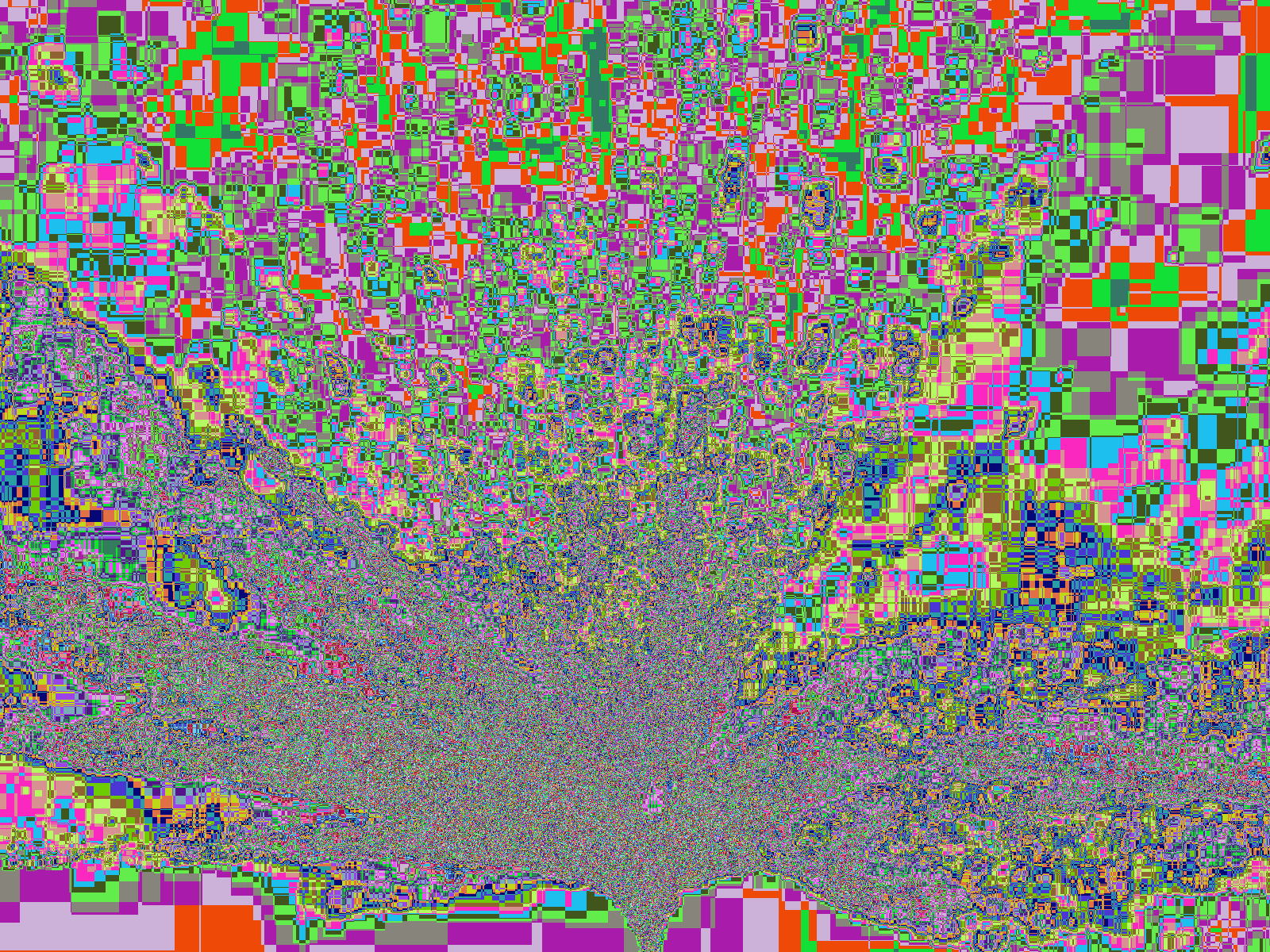}
\\
  {
    {\code{truck}}}
&
  {
    {\code{bmw}}}
&
  {
    {\code{powerplant}}}
&
  {
    {\code{CAD model}}}
&
  {
    {\code{PBRT landscape}}}
&
  {
    {\code{splatting}}}
\end{tabular}\\[-1em]
\end{centering}
\caption{\label{fig_scenes}Reference images for our test scenes; with
pseudo-color visualization of total number of triangles visited
along each ray (properly including even multiple hits at same distance).
From left to right: \code{truck} (426K triangles), PBRT \code{bmw} (385K triangles),
\code{powerplant} (12M triangles), a complete CAD \code{car} model (ca 28M triangles),
the PBRT \code{landscape} (30K instances, 4.3B triangles total),
and a set of triangles captured from a Gaussian Splatting
ray tracer (in this case, using all screen-aligned quads, ca 1.8M triangles).
}
\end{figure*}

\section{Front-to-Back Any-Hit}

The goal of this paper is to realize \emph{front to
back Any-Hit} (FTB): iterating through hits along a ray in guaranteed
front-to-back order, without missing hits even if there are 
multiple hits at same distance.  In this section, we will discuss
different ways of realizing this.

These will obviously differ in performance (we will look at that in
Section~\ref{sec:evaluation}), however, they can also differ in other
properties that an application may care about.  In particular, we
classify the different methods on the following properties:

\def\bla#1{\par\medskip\noindent\textbf{#1}}

\bla{Iiteration vs callback.} Some of our methods will allow the user
to manage his or her own iteration over the results by explicitly
asking, in each iteration, for the respectively next hit; with the
user being able to do arbitrary other work in between two such
calls. Other methods will only allow for a callback style operation
where the user provides some \code{userCode()} callback function, but
it is the FTB kernel (not the user) that controls when this function
is being called. In this case we assume the \code{userCode} can return
a value that indicates whether it does or does not want more
iterations to occur. To make our methods more comparable we will
formulate even the explicit iteration kernels over an explicit
\code{userCode()} function, but this formulation is not a requirement.
  
\bla{Access to RTX pipeline state} determines whether the
\code{userCode()} can access OptiX functions like
\code{optixTransformRay/Point/...()}, \code{optixGetPrimitiveIndex()},
etc.  This is assumed as given in traditional AH and CH programs, but
for our kernels is not: for example, for any kernel that does the
\emph{ignore-but-accept} discussed in
Section~\ref{sec:ignore-but-accept} the actually computed next hit
will \emph{not} be visible to the CH program, so any \code{userCode()}
would not have access to this data even if it were to be called in the
CH program.

For methods with access to RTX state the \code{userCode()} can
simply call \optix-getter functions as desired; for those that do not
the user has to modify the psuedo-code to store whatever
data he or she might need when the hit gets stored.

\bla{Stability wrt tree rebuild.} For some of our methods, the order
in which we encounter hits determines the order in which an AH program
encounters the respective triangles. This in turn may depend on the
order in which different triangles are stored in the respective BVH,
and depending on whether the BVH builder used to build this BVH is
temporally stable or not (ie, whether it always builds exactly the
same BVH or not) this may result in different order. This may or may
not matter to the application, but we do point out which methods are
vs are not stable with respect to BVHes built at different times.

A summary of our different methods with respect to
these properties is given in Table~\ref{tab-properties}.

\subsection{Method 1 (correct): Stable Order Next-Hit (\stableOrder)}

The root cause of why we cannot simply iterate through all hits
 with repeated CH traces is that in the RTX pipeline the only way
to specify that any hits should be ``behind'' what one already had is
to  specify \code{ray.tMin}---but this only tests for distance,
and thus cannot differentiate different hits with \emph{same} distance.
 
To fix this we first need to define a way of comparaing any two hits
in a way that, in math terms, forms a \emph{strict total
order}.
Once we have that, we can argue about which
hit A should be reported ``before'' another hit B even if A and B both
have the same distance. To do this, we can look at the triple of 
primitive index (\code{optixGetPrimitiveIndex()}), SBT index
(\code{optixGetSBTOffset()}), and instance index
(\code{optixGetInstanceIndex()}). These three values can
describe any possible intersection, so can be used to disambiguate any
two hits with same distance:
\begin{lstlisting}
// helper tuple that uniquely describes any hit
struct HitDesc {
  float t; int prim, geom, inst;
}

// comparator that allow for non-ambigous
// ordering of hits
bool operator<(HitDesc a, HitDesc b) {
  if (a.t != b.t) return a.t < b.t;
  if (a.inst != b.inst) return a.inst < b.inst;
  if (a.geom != b.geom) return a.geom < b.geom;
  if (a.prim != b.prim) return a.prim < b.prim;
  return false
}
\end{lstlisting}

Using this comparator we have a well defined way of describing what
the ``next'' hit is supposed to be: it has to be larger than the last
iteration's hit, and smaller than any other. We now craft an AH program
that checks all reported intersections for this property, and keeps
track of the hit that fulfills that condition. To evaluate this
condition the AH program also needs to know the hit reported by the
last iteration, and some storage to track its currently closest
result; both of these we simply store in some helper PRD (per ray
data) struct that we pass with the ray.

The core idea of this method should be fairly straightforward;
however, two caveats have to be considered. One is that the
pipeline already compares whether \code{ray.tMin < tHit}, so if the
last iteration's hit was at \code{tPrev} we can \emph{not} use that
as the next \code{ray.tMin}, and have to instead use
\code{nextBefore(tPrev)}, or other hits at that distance wouldn't even
show up in our AH program. The second caveat is that even if the AH
program does find a hit closer than what it has found so far---and
``accepts'' that in its own PRD---it \emph{still} has to tell the
pipeline to \emph{reject} that hit if it is at the same distance as the
closest one we have found so far. Conversely, any hit with a distance
strictly greater than what we have stored so far \emph{should} get
(\optix-)accepted so as to shorten the ray traversal;
even though we will obviouly not accept that hit itself. This means we
OptiX-accept some rays that we internally reject, and \optix-reject
others we actually store; this is confusing, but correct.

In pseudo-code, this method looks like this:

\begin{lstlisting}
// stable next-hit
struct StableNextHitPRD {
  // any hit has to be strictly '>' hitMin
  HitDesc hitMin;
  // tracks currently closest found hit.
  // any hit < this that we encounter during
  // traversal will overwrite this
  HitDesc hitMax;
};

// AH prog that finds actual next hit according
// to previously described '<' operator
void __AH() {
  HitDesc curr = 
  { optixTMax(), optixGetPrimitiveIndex(),... };
  auto &prd = getPRD<StableNextHitPRD>();
  if (prd.hitMin < curr && curr < prd.hitMax)
    prd.hitMax = curr;
    // caveat 2 (see text): do _not_ accept hits
    // at closest t, or rtx pipeline will 
    // automatically cull any others at this t
  if (curr.t == prd.hitMax.t)
    optixIgnoreIntersection();  
}
void front_to_back(Ray ray, UserPRD &userPRD) {
  float user_tMax = ray.tMax;
  StableNextHitPRD prd;
  prd.hitMin = { ray.tMin, -1, -1, -1 };
  while(true) {
    prd.hitMax = { user_tMax, -1, -1, -1 };
    optixTrace(ray, prd=&prd, DISABLE_CH);
    if (prd.tMax.primID < 0)
      break; // no hit found
    // user code has to be called explicitly, 
    // becauseAH prog (had to) optix-reject 
    userCode(prd.hitMax,userPRD);
    if (user code wants to exit)
      return
    // ... and iterate
    prd.hitMin = prd.hitMax;
    ray.tMin = justBelow(prd.hitMin.t)
    ray.tMax = user_tMax
  }
}
\end{lstlisting}

This method is temporally stable in that the order in which the AH
program visits triangles does not matter. It will also allow for an
iterative approach, and is not limited to callbacks. However, since
we have to \optix-reject hits that we actually accept the finally found
next hit will not be reported in a CH program, and will thus not have
access to the RTX state. Any state data the user needs will thus have to
be saved whenever the AH program (internally) accepts a new hit.

\subsection{Method 2: Reject-Repeats}

Method 1 worked by using a \code{operator<} to disambiguate hits at
same distance. An alternative way of doing so is to use the (temporal)
order in which they get reported by the RTX pipeline: Ie, assuming we
have already reported $N$ hits at a given distance \code{tPrev} we can
simply tell an AH program to simply skip the first $N$ hits at this
distance, and re-trace the previous ray, thus getting either the
N-plus-first hit at that distance, or one at a greater distance.

This is, in fact, the first method we used, but making it correct is
more tricky than it at first appears: The problem is that once we find
a hit at a given distance \code{tPrev} we want to make sure that the
next time we trace a ray we will use a \code{ray.tMin} of
\code{nextBefore(tPrev)}---but changing a ray's tMin value can---in
admittedly rare but still possible cases---change the order in which
the traversal visits leaves, and thus triangles.  Thus, the first hit
visited by the ray with modified \code{tMin} value may \emph{not} be
the same triangle that the predecessor found at that distance so
simply counting hits would be wrong.

This can be fixed by having the PRD store not only the reject counter,
but also store the last found hit; and the AH program to first test
for wether the current hit is the same as the last, and only use the
reject counter for all other hits at that distance. This sounds---and
is---tricky, but once done, is correct. Pseudo-code for this looks
like this:
\begin{lstlisting}
// reject-repeats
struct HelperPRD {
  HitDesc skipHit;
  int     skipCount;
  HitDesc thisHit;
};

void __AH() {
  auto &prd = getPRD<HelperPRD>();
  HitDesc thisHit = getCurrentHit();
  if (thisHit.dist > prd.skipDist)
    { /* accept this */ return; }
    // we know it cannot be smaller, so we're 
    // now at skipDist
  if (thisHit == prd.skipHit) {
    // if same as first at skip dist we
    // always ignore
    { optixIgnoreIntersection(); return; }
  if (prd.skipInfo.numLeftToSkip == 0)
    { /* no more skipping - accept! */ return; }
  // skip this one
  -- prd.skipInfo.numLeftToSkip;
  optixIgnoreIntersection();
}

void __CH() {
  // this method does let userCode()
  // have access to PRD
  auto ret = userCode();
  if (ret == keep_going) {
    // save thisHit for main loop logic
    prd.thisHit = getCurrentHit();
  }
}

void frontToBack(Ray ray) {
  float saved_tmax = ray.tmax;
  HelperPRD prd;
  prd.skipHit.dist = -inf;
  float next_tmin = ray.tmin;
  int   next_skipCount = 0;
  while (true) {
    ray.tmin = next_tmin;
    ray.tmax = saved_tmax;
    prd.thisHit.primID = -1; // mark ray as 'no hit'
    prd.skipCount = next_skipCount;
    traceRay(ray, PRD = &prd);
    if (ray.thisHit.primID < 0)
      // no hit found, or usercode wants to exit
      break;
    if (ray.thisHit.dist > prd.skipHit.dist) {
      // we found a new distance, (re-)init skipping
      next_tmin = justBelow(ray.thisHit.dist)
      // 0 _in addition to_ 'skipHit'!
      next_skipCount = 0;
      prd.skipHit = prd.thisHit;
    } else {
      // skip one more in next iteration
      ++nextSkipCount;
    }
  }
}
\end{lstlisting}

This method is not stable with respect to tree rebuilds because it
relies on the order in which the AH program visits hits. It does,
however, allow to be used in an iterative fashion, though because it
has to \optix-reject the true hit it will also not allow for the
\code{userCode()} to access the RTX state data.

\subsection{Method 3 (correct): \whileWhile}

The core idea of this kernel is that we can split our problem in two:
iterating over increasing unique hit \emph{distances} first, and then,
for each such distance, iterating over all hits that have that
distance.

For the first task, we  loop over closest hit rays
that always trace the next ray with a \code{tMin} set to whatever the
last found distance was. These rays are traced with AH disabled, and
with a CH program that does nothing other than returning the hit
distance through its per-ray data. We call these rays \emph{distance
feelers}, because all they do is find the next distance \emph{where}
something will happen.

Once we have a given distance \code{tNext}, we can then create a
second ray---that we call the ``executor'' ray---whose \code{tMin} we
set to \code{tMin=nextBefore(tHit)}, and its \code{tMax} to
\code{tMax=justAbove(tHit)}. We also equip this ray with an AH program
that executes the user's \code{userCode()} (hence the name), and then
\optix-rejects the found found hit.  Since there is only a single
possible float value inside an executor ray's \code{tMin/tMax}
interval (namely, \code{tHit}) that ray is guaranteed to have its AH
program called only for intersections at \code{tHit}, and since it
always \optix-rejects all rays it is also guaranteed to get called for
all hits at that distance. Executor do not need a CH program, so we
can actually use the same SBT entry for both types of rays: the SBT
entry holds both CH program (for feeler rays) and AH program (for
executors), and the trace call selects which one to use by
disabling the other.

Since this method effectively realizes two nested while loops---one
explicit one over feeler rays, then an implicit one over all AH
executions---we call this the \texttt{while-closest-while-any} method (or
\whileWhile, for short).  In pseudo-code, it looks like this:
\begin{lstlisting}
// while-closest-while-any
  
// helper CH prog that only returns hit dist
void __outerRay_CH() {
  float &prd = getPRD<float>();
  prd = optixTMax()
}

// AH prog for actual user code
void __innerRay_AH() {
  userCode();
  if (user code wants to terminate)
    optixTerminateRay();
  else
    optixIgnoreIntersection();
}
  
void front_to_back(Ray ray, UserPRD &userPRD) {
  // save original ray.tMax:
  float user_tMax = ray.tMax
  while(true) {
    float tNext = -1.f; // sentinel value
    optixTrace(ray, PRD=&tNext, DISABLE_AH)
    if (tNext == -1.f)
      // no next hit distance was found - done
      return;

    // restrict ray interval to ONLY tNext
    ray.tMin = justBelow(tNext)
    ray.tMax = justAbove(tNext)
    // trace ray with only AH program:
    optixTrace(ray, PRD=&userPRD, DISABLE_CH)

    // set next outer ray's interval to
    // ``anything larger than tNext''
    ray.tMin = tNext
    ray.tMax = user_tMax;
    // ... and iterate
  }
}
\end{lstlisting}

The downside of this method is that many intersections require
\emph{two} rays to be traced: one feeler ray for the distance, and
then a second one to execute. This sounds horrible, but actually
isn't: feeler rays are cheap because they have no AH program at all,
and a very cheap CH program---which makes them about as close to ideal
for the hardware as they could be. Executor rays do of course need an
AH program, and possibly multiple AH call; but with the extremely
narrow ray interval they require very(!) little BVH traversal and
triangle intersection, and execute the (costly!) AH programs only
where actually required.

Unlike \stableOrder this method can only be used in a
AH-program like callback fashion; the user code can terminate the
traversal, but cannot perform arbitrary operations between two
successive operations. On the other hand, user code is called exactly
when the pipeline found the respective hit, so the user code has full
access to the entire RTX pipeline state.

\subsection{Method 4 (correct): \whileMerged }

One way of avoiding the \whileWhile method's duplicate trace calls is
to merge the two kinds of rays---feelers and executors---into
one: The \whileWhile method uses only CH for one, and only AH
for the other; these can be combined as long as
the AH program gets called only on the same distance that the
\whileWhile method's AH programs would have been called, too.

To do this, we user a helper PRD that both specifies the last
iteration's found surface distance (\code{tExec}) as well as a slot
for storing the respectively next distance. We then change the AH
program to execute the user code at (only!) the intended distance, and
to also \optix-reject all hits at that distance. The CH program just
stores the hit distance \code{tFeeler}, as before, which becomes the
next iteration's executor distance.

During the iteration we  set the \code{tExec} to the previous
iteration's \code{tFeeler}, and start the ray with
\code{ray.tMin=justBelow(tExec)}. We call this the \whileMerged
kernel:
\begin{lstlisting}
// while-while-merged
struct HelperPRD {
  UserPRD *userPRD;
  float tExec;
  float tFeeler;
};

// helper CH prog that only returns hit dist
void __CH() {
  auto &prd = getPRD<HelperPRD>()
  // '-1' means 'user code wants to terminate
  if (prd.tFeeler != -2.f)
    prd.tFeeler = optixTMax()
}

// AH prog for actual user code
void __AH() {
  bool atExecDistance
    = (optixTMax()==prd.tExec);
  if (!atExecDistance)
    // not-first surf hits get accepted, but
    // do not call AH
    return;
  // first-surf hits call user code
  userCode();
  if (user code wants to terminate) {
    optixTerminateRay();
    prd.tFeeler = -2;
  } else
    optixIgnoreIntersection();
}
  
void front_to_back(Ray ray, UserPRD &userPRD)
{
  float user_tMax = ray.tMax
  HelperPRD prd { &userPRD, -1, -1 };
  while(true) {
    float tNext = -1.f; // sentinel value
    optixTrace(ray, PRD=&prd)
    if (prd.tFeeler < 0.f)
      // no surface dist was found - done
      return;

    // re-start right before next surface
    ray.tMin = justBelow(prd.tExec)
    ray.tMax = user_tMax;
    // enable AH for (only) that distance
    prd.tExec  = prd.tFeeler;
    prd.tFeeler = -1;
    // ... and iterate
  }
}
\end{lstlisting}

At first glance this should be a clear win over \whileWhile, because
it avoids that method's overhead of needing two rays for most hits.
However, the rays traced in this method have both AH and CH programs
enabled, have a non-singular \code{ray.tMin/tMax} ranges, and will
consequently call many more AH programs than the \whileWhile
variant. Even though these AH programs call the user code only where
required, they nevertheless make these rays much more expensive,
which---as our results will show---largely cancels out the savings
from fewer rays. Like the \whileWhile method this method
does allow the user code to access the RTX pipeline properties.

\subsection{For Reference: \ahOnly and \chOnly}

For reference, in the following evaluation of our methods we will also
compare to the two obvious (but incorrect) reference methods:
\chOnly runs a simple loop over successive calls to
\code{optixTrace} with each next ray's \code{tMin} set to the last
ray's found hit distance; with user code executed in the CH program,
and AH programs disabled.Similarly, \ahOnly traces a single ray with user code
called in the AH program, but no CH program. This method will
guarantee that no hits will get skipped, but might call the user code
out of proper distance order.

This methods will obviously not be correct; one will skip hits at
similar distance, the other reports hits out of order. Even as
performance baselines they have to be taken with a grain of
salt---e.g., finding only some hits will obviously be faster than properly finding
all---but we nevertheless decided to include these as well. We omit
sample code for this in
this paper; but reference code is available in the accompanying test
rig.

\begin{table*}[ht]
  \centering
\begin{tabular}{|l|ccc|}
  \hline
  method & correctness & type of & rtx state \\
         &             & iteration$^{(1)}$ & access$^{(2)}$  \\
  \hline
  \stableOrder & correct & explicit (return value) & no  \\
  \whileWhile  & correct & callback (in AH prog) & yes \\
  \whileMerged & correct & callback (in AH prog) & yes \\
  \rejectRepeats & correct & explicit (return value or CH prog) & yes \\
  \hline
  stable multi-hit variants & correct & explicit (return value) & no \\
  \hline
  \ahOnly & out of order  & callback (in AH prog) & yes \\
  \chOnly & skip coplanar  & explicit (return value or in CH prog)   & yes \\
  \hline
\end{tabular}
\caption{
  Summary of the different methods' properties.
  $^{(1)}$Type of iteration: \emph{explicit} means the user's code
  will explicitly ask for the next hit, and can do arbitrary other
  work between iterations; \emph{callback} means the user can only
  provide a callback function to be called (similar to an AH program),
  but once the trace is started the method itself determines when it
  gets called.  $^{(2)}$Whether the user code has access to the full
  RTX state (hit IDs, transforms, etc) through OptiX query
  functions.
\label{tab-properties}
}
\end{table*}

\begin{table*}[ht]
\centering
    {
      {
\begin{tabular}{c|cccccc}
\hline
method & Butler Truck & UNC PowerPlant & PBRT BMW & CAD model & PBRT Landscape & splatting\\
\hline
\hline
\multicolumn{7}{c}{maxDepth=1; always exist after first hit}\\
\hline
stable order & 0.13 {{(+1\%)}} & 0.13 {{(+7\%)}} & 0.18 {\color{red}(\textbf{+23\%)}} & 0.37 {\color{red}(\textbf{+62\%)}} & 1.01 {\color{red}(\textbf{+61\%)}} & 0.32 {\color{red}(\textbf{+70\%)}}\\
while-while & 0.13 {\color{green}(best)} & 0.12 {\color{green}(best)} & 0.15 {{(+6\%)}} & 0.25 {{(+10\%)}} & 0.69 {{(+10\%)}} & 0.20 {{(+4\%)}}\\
while-merged & 0.16 {\color{red}(\textbf{+19\%)}} & 0.16 {\color{red}(\textbf{+33\%)}} & 0.18 {\color{red}(\textbf{+23\%)}} & 0.38 {\color{red}(\textbf{+68\%)}} & 0.99 {\color{red}(\textbf{+58\%)}} & 0.30 {\color{red}(\textbf{+61\%)}}\\
reject-repeats & 0.13 {{(+0\%)}} & 0.12 {{(+1\%)}} & 0.15 {\color{green}(best)} & 0.23 {\color{green}(best)} & 0.63 {\color{green}(best)} & 0.19 {\color{green}(best)}\\
\hline
\hline
\multicolumn{7}{c}{maxDepth=5; iterate through first 5 hits (or all if less)}\\
\hline
stable order & 0.52 {\color{red}(\textbf{+47\%)}} & 0.49 {\color{red}(\textbf{+51\%)}} & 0.72 {\color{red}(\textbf{+34\%)}} & 1.70 {\color{red}(\textbf{+47\%)}} & 6.13 {\color{red}(\textbf{+68\%)}} & 1.74 {\color{red}(\textbf{+73\%)}}\\
while-while & 0.35 {\color{green}(best)} & 0.32 {\color{green}(best)} & 0.57 {{(+6\%)}} & 1.32 {\color{red}(\textbf{+14\%)}} & 3.72 {{(+2\%)}} & 1.00 {\color{green}(best)}\\
while-merged & 0.36 {{(+4\%)}} & 0.39 {\color{red}(\textbf{+20\%)}} & 0.57 {{(+6\%)}} & 1.28 {\color{red}(\textbf{+11\%)}} & 4.12 {\color{red}(\textbf{+13\%)}} & 1.26 {\color{red}(\textbf{+26\%)}}\\
reject-repeats & 0.35 {{(+1\%)}} & 0.36 {\color{red}(\textbf{+12\%)}} & 0.54 {\color{green}(best)} & 1.16 {\color{green}(best)} & 3.64 {\color{green}(best)} & 1.14 {\color{red}(\textbf{+14\%)}}\\
\hline
\hline
\multicolumn{7}{c}{maxDepth=25; iterate through first 25 hits (or all if less)}\\
\hline
stable order & 1.62 {\color{red}(\textbf{+85\%)}} & 1.79 {\color{red}(\textbf{+90\%)}} & 1.02 {\color{red}(\textbf{+32\%)}} & 5.50 {\color{red}(\textbf{+42\%)}} & 24.3 {\color{red}(\textbf{+64\%)}} & 5.33 {\color{red}(\textbf{+74\%)}}\\
while-while & 0.88 {\color{green}(best)} & 0.94 {\color{green}(best)} & 0.81 {{(+4\%)}} & 4.45 {\color{red}(\textbf{+15\%)}} & 15.3 {{(+3\%)}} & 3.06 {\color{green}(best)}\\
while-merged & 0.91 {{(+3\%)}} & 1.18 {\color{red}(\textbf{+25\%)}} & 0.78 {\color{green}(best)} & 3.88 {\color{green}(best)} & 14.9 {\color{green}(best)} & 3.56 {\color{red}(\textbf{+17\%)}}\\
reject-repeats & 1.05 {\color{red}(\textbf{+20\%)}} & 1.13 {\color{red}(\textbf{+20\%)}} & 0.78 {{(+1\%)}} & 3.89 {{(+0\%)}} & 14.9 {{(+0\%)}} & 3.53 {\color{red}(\textbf{+15\%)}}\\
\hline
\hline
\multicolumn{7}{c}{no max depth; iterate through all hits, no matter how many}\\
\hline
stable order & 2.18 {\color{red}(\textbf{+106\%)}} & 3.09 {\color{red}(\textbf{+60\%)}} & 1.03 {\color{red}(\textbf{+32\%)}} & 6.95 {\color{red}(\textbf{+50\%)}} & 59.4 {\color{red}(\textbf{+65\%)}} & 40.5 {\color{red}(\textbf{+144\%)}}\\
while-while & 1.06 {\color{green}(best)} & 1.92 {\color{green}(best)} & 0.81 {{(+4\%)}} & 5.23 {\color{red}(\textbf{+13\%)}} & 36.1 {\color{green}(best)} & 16.6 {\color{green}(best)}\\
while-merged & 1.15 {{(+8\%)}} & 2.09 {{(+9\%)}} & 0.78 {\color{green}(best)} & 4.62 {\color{green}(best)} & 36.7 {{(+2\%)}} & 23.5 {\color{red}(\textbf{+42\%)}}\\
reject-repeats & 1.37 {\color{red}(\textbf{+28\%)}} & 2.11 {{(+10\%)}} & 0.79 {{(+1\%)}} & 4.70 {{(+2\%)}} & 37.5 {{(+4\%)}} & 23.6 {\color{red}(\textbf{+42\%)}}\\
\hline
\hline
\multicolumn{7}{c}{probabilistic depth; each iteration stops with 25\% chance}\\
\hline
stable order & 0.59 {\color{red}(\textbf{+72\%)}} & 0.64 {\color{red}(\textbf{+49\%)}} & 0.58 {\color{red}(\textbf{+28\%)}} & 1.97 {\color{red}(\textbf{+45\%)}} & 6.94 {\color{red}(\textbf{+70\%)}} & 1.60 {\color{red}(\textbf{+69\%)}}\\
while-while & 0.34 {\color{green}(best)} & 0.43 {\color{green}(best)} & 0.46 {{(+1\%)}} & 1.42 {{(+4\%)}} & 4.09 {\color{green}(best)} & 0.94 {\color{green}(best)}\\
while-merged & 0.38 {{(+10\%)}} & 0.49 {\color{red}(\textbf{+12\%)}} & 0.47 {{(+4\%)}} & 1.42 {{(+4\%)}} & 4.54 {\color{red}(\textbf{+11\%)}} & 1.16 {\color{red}(\textbf{+23\%)}}\\
reject-repeats & 0.41 {\color{red}(\textbf{+19\%)}} & 0.49 {\color{red}(\textbf{+14\%)}} & 0.46 {\color{green}(best)} & 1.36 {\color{green}(best)} & 4.28 {{(+5\%)}} & 1.10 {\color{red}(\textbf{+17\%)}}\\
\end{tabular}
}}
\caption{Performance for the different kernels that iterate over hits individually, in a hit-by-hit fashion, for different scenes and different patterns of how many iterations the application will do. Performance numbers are in milliseconds per $1600\times 1200$ frame, on a NVIDIA RTX 4090 GPU, for iterating through the specified number of iterations with intentionally simpler per-intersection \texttt{userCode()}.\label{tab-perf-hit-by-hit} }
\end{table*}

\section{Evaluation}
\label{sec:evaluation}

Our kernels will not only differ in performance,
but also in other properties such as iterator vs callback, whether or
not it has access to the pipeline state, etc. A side by side
comparison regarding these properties is given in
Table~\ref{tab-properties}. Regarding performance, how the different
methods perform depends not only on the actual scene, but also on how
many iterations are to be performed.

\subsection{Comparison Methodology}

To evaluate all these options we have written a small test-rig that
can load a model (with instances of triangle meshes), generate primary
rays for what we consider a representative view of the respective
model, and then in each pixel iterate through multiple hits along each
rays. To compare the different method we use a framework where all
kernels operate in a callback fashion where the actual user code to be
executed is a function that operates on a user-defined PRD struct, and
returns a bool that indicates whether it does or doesn't want to
terminate this ray. For each method we can then compile OptiX device
code that calls this same user code in whatever fashion this kernel
operates in.

To compare how different kernels react to different iteration counts
we use multiple different such user-defined callbacks: one counts how
often it got called, and terminates traversal after a compile-time
defined maximum depth $N$; the other generates a pseudo-random number
generator for each ray, and uses that to probabilistically terminates
rays with a certain probability of 1-in-N (i.e., for
\code{probDepth=4} the userCode terminates a ray with a 25\%
chance). The latter of these kernels is particularly useful to
simulate situations like rays going through foliage with partially
transparent surfaces; which will act largely proabilistically. The
kernel with a fixed maximum depth is useful in evaluating both
extremes of actually stopping after the first hit (maxDepth=1) as well
as going through all hits that exist (maxDepth=inf); as well as for
some values in between.

To ensure correctness our user code tracks how often it has been
called, as well as the distance for which it was last
invoked. Combining this with different \code{probDepth} and
\code{maxDepth} variants allows for thorough validation:
methods that miss or double-report hits would
produce at least some different pixels for \code{maxDepth=none}, while
methods that didn't produce the proper order would produce different
depth-shaded results when called with a fixed \code{maxDepth}. We 
use a pseudo-color generator to make sure that even very small
differences in computed values will produce very different colors, so
differences cannot get masked by image compression.

We test on different test scenes (see
Figure~\ref{fig_scenes}). These are intentionally chosen to cover a
wide range of different models, including with and without instances,
man-made structures vs vegetation, etc. The \code{Butler Truck} was a typical
test case for the \emph{Bullet Ray Vision}~\cite{brv}
and similar multi-hit related projects, and \code{splatting}
represents the sort of screen-space layered quads one can encounter in
certain gaussian splatting based rendering. We also point out how
different the otherwise similar-looking \code{PBRT BMW} and \code{CAD
  Model} are for this evaluation: one contains only externally visible
cars (as could be expected for a typical rendering model), while the
other contains all the interior geometry as well (as can be expected
for a more engineering-related CAD model).

\begin{table*}[ht]
\centering
    {
      {
\begin{tabular}{c|cccccc}
\hline
method & truck & powerplant & PBRT BMW & CAD model & landscape & splatting\\
\hline
\hline
\multicolumn{7}{c}{maxDepth=1; always exist after first hit}\\
\hline
(best hit-by-hit) & 0.13 & 0.12 & 0.15 & 0.23 & 0.63 & 0.19\\
\hline
stable multiHit(4) & 0.19 {\color{red}(+48\%)} & 0.25 {\color{red}\textbf{(+2.0x)}} & 0.28 {\color{red}(+90\%)} & 0.62 {\color{red}\textbf{(+2.8x)}} & 2.19 {\color{red}\textbf{(+3.5x)}} & 0.58 {\color{red}\textbf{(+3.1x)}}\\
stable multiHit(16) & 0.37 {\color{red}\textbf{(+2.9x)}} & 0.51 {\color{red}\textbf{(+4.1x)}} & 0.35 {\color{red}\textbf{(+2.4x)}} & 1.19 {\color{red}\textbf{(+5.3x)}} & 5.48 {\color{red}\textbf{(+8.7x)}} & 1.53 {\color{red}\textbf{(+8.2x)}}\\
stable multiHit(64) & 0.57 {\color{red}\textbf{(+4.4x)}} & 1.26 {\color{red}\textbf{(+10.3x)}} & 0.36 {\color{red}\textbf{(+2.4x)}} & 2.12 {\color{red}\textbf{(+9.4x)}} & 13.3 {\color{red}\textbf{(+21.1x)}} & 5.91 {\color{red}\textbf{(+31.5x)}}\\
\hline
\hline
\multicolumn{7}{c}{maxDepth=5; iterate through first 5 hits (or all if less)}\\
\hline
(best hit-by-hit) & 0.35 & 0.32 & 0.54 & 1.16 & 3.64 & 1.00\\
\hline
stable multiHit(4) & 0.34 {\color{green}(--2\%)} & 0.42 {\color{red}(+31\%)} & 0.40 {\color{green}(--25\%)} & 1.14 {\color{green}(--2\%)} & 4.65 {\color{red}(+28\%)} & 1.10 {\color{red}(+9.8\%)}\\
stable multiHit(16) & 0.38 {\color{red}(+8.2\%)} & 0.53 {\color{red}(+64\%)} & 0.35 {\color{green}(--34\%)} & 1.20 {\color{red}(+3.9\%)} & 5.48 {\color{red}(+50\%)} & 1.54 {\color{red}(+54\%)}\\
stable multiHit(64) & 0.57 {\color{red}(+64\%)} & 1.28 {\color{red}\textbf{(+4.0x)}} & 0.36 {\color{green}(--33\%)} & 2.14 {\color{red}(+84\%)} & 13.3 {\color{red}\textbf{(+3.6x)}} & 5.95 {\color{red}\textbf{(+5.9x)}}\\
\hline
\hline
\multicolumn{7}{c}{maxDepth=25; iterate through first 25 hits (or all if less)}\\
\hline
(best hit-by-hit) & 0.88 & 0.94 & 0.78 & 3.88 & 14.9 & 3.06\\
\hline
stable multiHit(4) & 0.73 {\color{green}(--16\%)} & 1.07 {\color{red}(+14\%)} & 0.47 {\color{green}(--40\%)} & 2.41 {\color{green}(--38\%)} & 12.4 {\color{green}(--16\%)} & 2.39 {\color{green}(--22\%)}\\
stable multiHit(16) & 0.52 {\color{green}(--40\%)} & 0.83 {\color{green}(--12\%)} & 0.37 {\color{green}(--52\%)} & 1.68 {\color{green}(--57\%)} & 8.90 {\color{green}(--40\%)} & 2.54 {\color{green}(--17\%)}\\
stable multiHit(64) & 0.58 {\color{green}(--34\%)} & 1.28 {\color{red}(+36\%)} & 0.37 {\color{green}(--52\%)} & 2.15 {\color{green}(--45\%)} & 13.2 {\color{green}(--11\%)} & 6.00 {\color{red}(+96\%)}\\
\hline
\hline
\multicolumn{7}{c}{no max depth; iterate through all hits, no matter how many}\\
\hline
(best hit-by-hit) & 1.06 & 1.92 & 0.78 & 4.62 & 36.1 & 16.6\\
\hline
stable multiHit(4) & 0.91 {\color{green}(--14\%)} & 1.61 {\color{green}(--16\%)} & 0.47 {\color{green}(--40\%)} & 3.11 {\color{green}(--33\%)} & 27.8 {\color{green}(--23\%)} & 18.4 {\color{red}(+11\%)}\\
stable multiHit(16) & 0.61 {\color{green}(--42\%)} & 1.19 {\color{green}(--38\%)} & 0.36 {\color{green}(--54\%)} & 2.04 {\color{green}(--56\%)} & 18.8 {\color{green}(--48\%)} & 15.9 {\color{green}(--5\%)}\\
stable multiHit(64) & 0.57 {\color{green}(--46\%)} & 1.54 {\color{green}(--20\%)} & 0.36 {\color{green}(--54\%)} & 2.26 {\color{green}(--51\%)} & 19.3 {\color{green}(--47\%)} & 21.1 {\color{red}(+27\%)}\\
\hline
\hline
\multicolumn{7}{c}{probabilistic depth; each iteration stops with 25\% chance}\\
\hline
(best hit-by-hit) & 0.34 & 0.43 & 0.46 & 1.36 & 4.09 & 0.94\\
\hline
stable multiHit(4) & 0.35 {\color{red}(+2.7\%)} & 0.48 {\color{red}(+11\%)} & 0.36 {\color{green}(--21\%)} & 1.12 {\color{green}(--18\%)} & 4.04 {\color{green}(--1\%)} & 0.97 {\color{red}(+3.4\%)}\\
stable multiHit(16) & 0.41 {\color{red}(+19\%)} & 0.58 {\color{red}(+34\%)} & 0.38 {\color{green}(--17\%)} & 1.30 {\color{green}(--5\%)} & 5.66 {\color{red}(+38\%)} & 1.66 {\color{red}(+76\%)}\\
stable multiHit(64) & 0.61 {\color{red}(+77\%)} & 1.31 {\color{red}\textbf{(+3.0x)}} & 0.39 {\color{green}(--15\%)} & 2.22 {\color{red}(+63\%)} & 13.2 {\color{red}\textbf{(+3.2x)}} & 6.05 {\color{red}\textbf{(+6.4x)}}\\
\end{tabular}
}}
\caption{Time per frame (in ms, $1600\times 1200$ pixels, on a NVIDIA RTX~4090 GPU) per frame for the different multi-hit methods relative to the respectively best hit-by-hit method, for various maximum number of iterations. Green is better, red is worse. \label{tab-multi-hit} }
\end{table*}

\subsection{Comparison of Individual Hit Methods}

As we will show below the multi-hit methods will have---depending on
chosen iteration count---vastly different performance characteristics
than those methods that iterate on a hit by hit basis.

To avoid this from masking how the other methods perform we will first
only compare those kernels that operate on an individual hit
basis. Performance results for different test-rig configuration and
scenes are given in Table~\ref{tab-perf-hit-by-hit}.
This table shows that there is no clear winner that is best in all cases. However, a few observations can still be
made. On is that \stableOrder is never good. It lacks access to the pipeline
state, and always performs poorly: each iteration requires a new ray to be
traced, each such ray has to use a non-trivial AH program, and those
AH programs cannot even accept rays at the
closest distance, causing many such calls. This method does
allow for explicit iteration, but so does \code{reject-repeats}, which
is almost always faster, while also allowing for access to the RTX
pipeline state. Together, this
leaves little reason to ever recommend this method---which is
concerning in that this was---before this evaluation---our go-to
method for this problem.

Among the remaining three methods, \code{while-while} is consistently
better than \code{while-merged}. This is because AH programs are
expensive, so despite tracing more rays \code{while-while}
still wins, because the it has fewer AH calls overall. \code{While-while}
is also easier to implement, and with all other properties the same
this leaves \code{while-while} a subtle but nevertheless clear winner
over \code{while-merged}.

This leaves \code{while-while} versus \code{reject repeats}. Which one
of those is faster depends on the scene; however, we believe this to
not be random, but instead, to depend on what fraction of the scene's
triangles will spatially overlap other triangles: The \code{Butler
  Truck} contains lots of co-planar surfaces, which obviously overlap;
the \code{powerplant} contains less fully co-planar triangles, but in its
interior it does contain lots of long diagonal triangles that a BVH
with only axis-aligned boxes probably cannot separate. In either of
these two cases, being able to iterate over multiple primitives in the
same BVH node with an AH program should be faster than iterating with
a new ray---but though we believe this to be plausible, we stress that
this is merely a hypothesis. Either way, neither of the two methods is
a clear winner over the other, and even where they differ they do so
by typically less than 20\%.

Outside of performance \code{while-while} is arguably slightly easier
to implement than \code{reject repeats}. Both have access to the RTX
pipeline stack, but the former can only be realized inside an AH
callback framework, while the latter can be used in both callback form
as well as in a CH program.

\def\tableAdjust{}
\begin{table*}[ht]
\centering
    {
      {
\begin{tabular}{c|cccccc}
\hline
method & truck & powerplant & PBRT BMW & CAD model & landscape & splatting\\
\hline
\hline
\multicolumn{7}{c}{maxDepth=1; always exist after first hit}\\
\hline
(best hit-by-hit) & 0.13 & 0.12 & 0.15 & 0.23 & 0.63 & 0.19\\
\hline
reference AH only & 0.08 {\color{green}(-37\%)} & 0.08 {\color{green}(-34\%)} & 0.09 {\color{green}(-39\%)} & 0.12 {\color{green}(-46\%)} & 0.40 {\color{green}(-37\%)} & 0.10 {\color{green}(-48\%)}\\
reference CH only & 0.11 {\color{green}(-12\%)} & 0.11 {\color{green}(-7\%)} & 0.11 {\color{green}(-22\%)} & 0.18 {\color{green}(-20\%)} & 0.58 {\color{green}(-9\%)} & 0.13 {\color{green}(-31\%)}\\
\hline
\hline
\multicolumn{7}{c}{maxDepth=5; iterate through first 5 hits (or all if less)}\\
\hline
(best hit-by-hit) & 0.35 & 0.32 & 0.54 & 1.16 & 3.64 & 1.00\\
\hline
reference AH only & 0.14 {\color{green}(-60\%)} & 0.16 {\color{green}(-52\%)} & 0.25 {\color{green}(-55\%)} & 0.45 {\color{green}(-62\%)} & 1.46 {\color{green}(-60\%)} & 0.37 {\color{green}(-63\%)}\\
reference CH only & 0.24 {\color{green}(-30\%)} & 0.26 {\color{green}(-20\%)} & 0.34 {\color{green}(-38\%)} & 0.89 {\color{green}(-23\%)} & 2.88 {\color{green}(-21\%)} & 0.53 {\color{green}(-47\%)}\\
\hline
\hline
\multicolumn{7}{c}{maxDepth=25; iterate through first 25 hits (or all if less)}\\
\hline
(best hit-by-hit) & 0.88 & 0.94 & 0.78 & 3.88 & 14.9 & 3.06\\
\hline
reference AH only & 0.31 {\color{green}(-64\%)} & 0.40 {\color{green}(-57\%)} & 0.33 {\color{green}(-58\%)} & 1.11 {\color{green}(-71\%)} & 5.42 {\color{green}(-64\%)} & 0.96 {\color{green}(-69\%)}\\
reference CH only & 0.55 {\color{green}(-37\%)} & 0.74 {\color{green}(-22\%)} & 0.45 {\color{green}(-41\%)} & 2.82 {\color{green}(-27\%)} & 12.2 {\color{green}(-18\%)} & 1.62 {\color{green}(-47\%)}\\
\hline
\hline
\multicolumn{7}{c}{no max depth; iterate through all hits, no matter how many}\\
\hline
(best hit-by-hit) & 1.06 & 1.92 & 0.78 & 4.62 & 36.1 & 16.6\\
\hline
reference AH only & 0.37 {\color{green}(-65\%)} & 0.65 {\color{green}(-66\%)} & 0.33 {\color{green}(-58\%)} & 1.26 {\color{green}(-73\%)} & 9.27 {\color{green}(-74\%)} & 2.32 {\color{green}(-86\%)}\\
reference CH only & 0.63 {\color{green}(-41\%)} & 1.28 {\color{green}(-33\%)} & 0.46 {\color{green}(-42\%)} & 3.23 {\color{green}(-30\%)} & 25.5 {\color{green}(-29\%)} & 8.43 {\color{green}(-49\%)}\\
\hline
\hline
\multicolumn{7}{c}{probabilistic depth; each iteration stops with 25\% chance}\\
\hline
(best hit-by-hit) & 0.34 & 0.43 & 0.46 & 1.36 & 4.09 & 0.94\\
\hline
reference AH only & 0.14 {\color{green}(-59\%)} & 0.19 {\color{green}(-56\%)} & 0.20 {\color{green}(-55\%)} & 0.47 {\color{green}(-65\%)} & 1.56 {\color{green}(-62\%)} & 0.32 {\color{green}(-66\%)}\\
reference CH only & 0.25 {\color{green}(-27\%)} & 0.36 {\color{green}(-18\%)} & 0.29 {\color{green}(-37\%)} & 0.95 {\color{green}(-30\%)} & 3.12 {\color{green}(-24\%)} & 0.52 {\color{green}(-45\%)}\\
\end{tabular}
}}
\caption{Baseline performance comparison to the two (incorrect) reference methods. \tableAdjust \label{tab:perf-baseline} }
\end{table*}

\subsection{Comparison to Multi-Hit Methods}

In Table~\ref{tab-multi-hit} we also include multi-hit methods into
this comparison. To make this easier
we only compare those to the respectively best result from a
hit-by-hit method. The outcome of this experiment
(Table~\ref{tab-multi-hit}) is clear: The best way to view the
multi-hit methods is as some sort of prefetching scheme that uses a
single (and much more expensive) trace to find multiple hits at once,
and then amortizes the cost of that the next $N$ iterations.

Viewed this way, the outcome is obvious: For configuration that end up
iterating very often the multi-hit methods are much faster than
iterating hit by hit, and overall, the more hits we find in each
multi-hit call the better for performance. Conversely, configuration
that on average take fewer iterations will be much slower that for the
hit-by-hit methods, and the larger the number of hits we gather in the
multi-hit the worse it gets. For rather deep iteration counts these
methods can pay off, other they can be significantly worse.

There are also other considerations the prospective user
may want to take into account: on one hand, multi-hit methods provide
more information than hit-by-hit method, which may be useful if, for
example, the simulation code has to sort multiple same-distance hits
by other properties such as material type, or whether the ray this
front or back face of the triangle. On the other hand, multi-hit
methods cannot access the OptiX state, and require (much) more kernel
memory to store all these hit points, which might be prohibitive.

\subsection{Comparison to Reference Baseline}

In Table~\ref{tab:perf-baseline} we also
compare the respectively best correct hit-by-hit method to the
\ahOnly and \chOnly variants. This
comparison has to be taken with a grain of salt, as the latter
will almost always report only a subset of the correct hits.

Not surprisingly these reference methods are always faster.  For the
two simplest models (PBRT BMW and Butler Truck) the CH method is up to
almost 2x faster, but for the more interesting model these methods'
runtime is only about a quarter or a third lower. This in turns means
that our kernels are almost surprisingly close to what the hardware
can actually deliver, which is encouraging (where applicable the
multi-hit methods would even beat the CH reference methods).

The AH method is obviously even faster; often by about 2x faster than
our best method, and sometimes even close to 3x. If proper depth
ordering of the results is not required this is obviously
preferable. However, we do point out that our test images show that
this method will, indeed, typically report hits out of order.

\section{Summary, Discussion, and Conclusion}

In this paper, we have described and evaluated different methods for
solving what we call the \emph{front-to-back any-hit (FTB)} problem---the
problem of iterating through multiple hits along a ray, in
front-to-back sorted order, without skipping any hits even if they
occur at same distance. We presented five methods for
doing so in an iterative hit-by-hit fashion, as well as a multi-hit
variant that---unlike classical multi-hit methods---also allows for
always finding the respectively \emph{next} N hits. We have classified
the different methods according to various properties, and evaluated
their performance using a thorough test-rig that allows for simulating
different kinds of applications that require different numbers of
iterations and even probabilistic iteration depth.

We implemented all our kernels in OptiX~\cite{optix}, but realizations
in Vulkan~\cite{vkr-spec}, DirectX~\cite{dxr-spec}, Metal~\cite{metal}, or
even Embree~\cite{embree} or HipRT~\cite{HipRT} should be
straightforward. We believe our implementation to be reasonably
efficient; however, our test-rig and the requirement to run all
kernels in the same framework will almost certainly carry some
overhead, so a dedicated optimization effort for a single kernel may
well see some improvements. This is particularly true for the
multi-hit method, which is currently based on the same idea as
\stableOrder, which itself was generally slower than the
\rejectRepeats or \whileWhile. This suggests that basing
multi-hit on those methods might even faster multi-hit kernels---but
would probably not change the general findings of our evaluation.

A big issue with our evaluation is that all our experiments used a
very simple \code{userCode()} mock-up that only required very little
hit data. Real applications would likely have to track much more
information per hit---which may affect some kernels more than others,
and in particular might lead to significant differences between those
kernels that can access the RTX state and those that do not. Without
actual end-user applications, this is very hard to evaluate.

Ultimately what we would like to see is to have FTB added to the RTX
pipeline; either as a special trace call, or through a special flag
passed to the trace function. This could happen in the same was a
trace can today be told to ignore AH or CH calls, or to stop at the
first intersection, etc---except it would simply enforce AH calls to
be delivered in order. Since our kernels already operate only on
existing RTX infrastructure, doing so should be straightforward in the
driver, but this has not yet happened. Adopting FTB into the RTX spec
might also allow for additional optimizations inside the driver that
are not currently possible when having to use only publicly exposed
functionality, but how much that might gain is hard to tell.

\bibliographystyle{ACM-Reference-Format}
\bibliography{main}


\end{document}